\documentclass[conference]{IEEEtran}
\IEEEoverridecommandlockouts
\usepackage[numbers]{natbib}
\setcitestyle{open={[},close={]}}

\usepackage{amsmath,amssymb,amsfonts}

\usepackage{graphicx}
\usepackage{textcomp}
\usepackage{xcolor}
\usepackage{float}
\usepackage{array}
\usepackage[font=footnotesize,labelfont=bf]{caption}
\usepackage{optidef}
\usepackage{subcaption}
\usepackage[utf8]{inputenc}
\def\BibTeX{{\rm B\kern-.05em{\sc i\kern-.025em b}\kern-.08em
    T\kern-.1667em\lower.7ex\hbox{E}\kern-.125emX}}

\usepackage{algorithm}
\usepackage{algpseudocode}
\usepackage{color, xcolor, colortbl}

\begin{document}

\title{Electrical Load Forecasting Using Edge Computing and Federated Learning\\}

\author{\IEEEauthorblockN{
		Afaf Taïk  and
		Soumaya Cherkaoui }	
	\IEEEauthorblockA{
		INTERLAB, Engineering Faculty, Université de Sherbrooke, Canada.\\ 
		\{afaf.taik, soumaya.cherkaoui\}@usherbrooke.ca }}
\maketitle

\begin{abstract}
In the smart grid, huge amounts of consumption data are used to train deep learning models for applications such as load monitoring and demand response. However, these applications raise concerns regarding security and have high accuracy requirements. In one hand, the data used is privacy-sensitive. For instance, the fine-grained data collected by a smart meter at a consumer's home may reveal information on the appliances and thus the consumer's behaviour at home. On the other hand, the deep learning models require big data volumes with enough variety and to be trained adequately. In this paper, we evaluate the use of Edge computing and federated learning, a decentralized machine learning scheme that allows to increase the volume and diversity of data used to train the deep learning models without compromising privacy. This paper reports, to the best of our knowledge, the first  use of federated learning for household load forecasting and achieves promising results. The simulations were done using Tensorflow Federated on the data from 200 houses from Texas, USA.   
\end{abstract}

\IEEEpeerreviewmaketitle


\section{Introduction}
\label{sec:introduction}
Load forecasting is an essential part of the development of the smart grid. Long-term load forecasting is deemed necessary for infrastructure planning, while mid-term and short-term load forecasting are key tasks in system operations \cite{i1}.
Day-to-day operational efficiency of electrical power delivery, in particular, requires an accurate prediction of short-term load profiles, which is based on collecting and analysing large volumes of high-resolution data from households.  However, individual short-term load forecasting (STLF) has been proven to be a challenging task because of profile volatility. In fact, the electrical load of a house has a high correlation to its residents' behaviour, which is too stochastic and often hard to predict \cite{r5,said_advanced_2013}.

Benchmarks for state-of-the-art methods \cite{r2,r3} have found that deep neural networks are a promising solution for the STLF problem at the household level, due to their ability to capture complex and non-linear patterns. Neural networks outperform other prediction methods such as Auto Regressive Integrated Moving Average (ARIMA)\cite{filali_prediction-based_2019} and Support Vector Regression (SVR). Nevertheless, applying deep learning models alone will not lead to significant improvements, as models tend to suffer from overfitting \cite{i4}. An overfitted model is a model that learned the details of the training data including the noise, which affects its ability to generalize when applied to new data. To tackle this issue, it is recommended to increase the diversity and size of the used data by combining
usage records from different households. Typically, proposed frameworks \cite{r6,r7} assume that all data records are transferred from smart meters to a centralized computational infrastructure through broadband networks to train models. Nevertheless, this assumption raises concerns related to privacy, since the load profiles reveal a lot of sensitive information, such as device usage and the household’s occupancy. Sending such detailed data over networks makes it exposed to malicious interception and misuse.  \\
To address privacy concerns while still increasing data records’ volume and variety, a new on-device solution was recently proposed by the Machine Learning community: Federated Learning (FL) \cite{f3}. Federated Learning is a decentralized machine learning scheme, where each device participates in training a central model without sending any data. As illustrated in Fig.1, the server first initializes the model either arbitrarily or by using publicly available data. Then, the model is sent to a set of randomly selected devices (clients) for local training using their data. Each client sends to the server an update of the model's weights, which will be averaged and used to update the global model. This process will be repeated until the global model stabilizes.

\begin{figure}[t]
	\centering
	\includegraphics[scale=0.50]{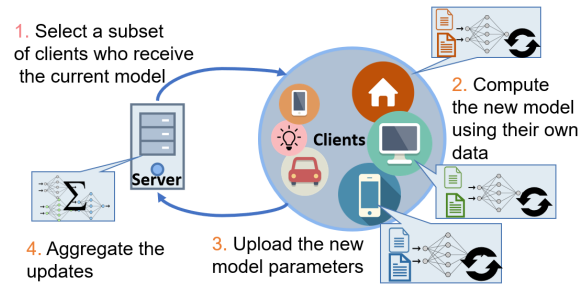}
	\caption{Iterative communications between clients and server in Federated Learning}
	\label{fig:fig_distribution}
\end{figure}

The main purpose of this paper is to evaluate the use of Edge computing, together with the Federated Learning approach in the STLF challenge for electricity in households. Edge computing refers to data processing  at the edge of a network as opposed to  cloud or remote server processing. We use Long-short Term Memory (LSTM) \cite{lstm1}, a deep neural network for forecasting time series, which uses previous observations of the house’s electrical load to predict future ones. We study a group of houses that have similar properties (geographical location, type of building), on a short period of time to avoid the weather's fluctuations and seasonality impact. Federated learning is performed on houses grid Edge equipment. Edge equipment is usually present at the end of the electrical distribution system as a smart interface between the customer and the electric power supply, be it a smart meter or a more sophisticated equipment. Our contributions in this work can be summarized as follows: (1) We propose an enabling architecture for FL using Edge equipment in the smart grid; (2)  We evaluate the potential gain of FL in terms of accuracy through simulations; and (3) we evaluate the potential network load gain through numerical results. To these contributions, we add the gain in privacy leveraged by decentralization and Edge computing.

The remainder of this paper is structured as follows: Section II discusses related works focusing on load prediction and privacy. In Section III, we define the proposed approach and used methods. Section IV introduces the simulations and numerical results. Then in Section V we discuss the limitations and future work. Section VI concludes the paper.

\section{Related work}
\label{Related work}

Many recent research works used deep neural networks, and particularly Long-short term memory (LSTM) to tackle the short-term load forecasting challenge. In fact, benchmarks have proved LSTM's potential  compared to other methods\cite{i2,i3}, yet the results do not match the level of desired exactitude in terms of Root Mean Square Error (RMSE) and Mean Average Percentage Error (MAPE). In order to improve forecasting accuracy, authors in \cite{r1} propose to use a variant of LSTM that is a sequence-to-sequence LSTM, which gives better results for one-minute resolution data, but no significant improvement for the one-hour resolution compared to standard LSTM. Furthermore, other authors \cite{r2} consider the problem of finding the best LSTM network to be a hyperparameter tuning problem, and use the genetic algorithm to this end. They state that finding the best combination of window size and number of hidden neurons in each layer remains a probabilistic task. 

Some other works see that the problem is not simply an neural network architecture problem, and that ability of generalization of data-driven forecasting models is the real issue. In fact, many of the proposed models' accuracy drops when they are applied to new datasets \cite{r3}. Some works suggest to use complementary data about the weather \cite{r4} or records from the appliances \cite{r5}. While the weather has a real impact on the aggregated electrical consumption, the individual short-term load is more related to the occupants' behaviour\cite{said_advanced_2013,said_scheduling_2014,rezgui_smart_2017}. However, collecting data from appliances around each house is an expensive and privacy-intrusive task.

Another approach to enrich the training data is grouping data from several customers. Authors in \cite{r6} use clustering to group users with similar profiles, hence reducing the variance of uncertainty within groups. Authors in \cite{r7} propose a pooling technique that increases data's diversity to overcome the overfitting problem. Nonetheless, these methods are heavily centralized and are prone to privacy-issues.   

Fine-grained consumption data sent over networks is subject to many privacy threats when leaked through unauthorized interception or eavesdropping \cite{r8}. Many efforts were conducted to protect the users' identities in the smart grid. For instance, authors in \cite{r9} propose a clustering-based method where each group of users who are geographically close receive a common serial number. However this method makes it hard to treat each client individually because of the anonymity. Other works' focus is masking the consumption data, where data aggregation is the most popular method \cite{r10,r11}, but it goes in opposite directions with STLF requirements.       

 In regards to user privacy and prediction accuracy, none of the aforementioned papers address both of these aspects. In the proposed work, we suggest to use the Edge Equipment that compose the Home Area Network (HAN) to carry out operations related to client selection and training neural network at the Edge following the federated learning scheme, allowing the use of data to train a global model without compromising the resident's privacy.

\section{System Model}
\label{sec:used methods}
We propose the network architecture shown in Fig.2 with two main components: a Multi-access Edge Computing (MEC) server \cite{mec}  and clients. Clients are houses with Edge equipment which is essentially composed of smart-meters and other devices in the HAN. FL is used to build a global LSTM-based model for STLF. The training rounds are orchestrated by the MEC server and executed by the clients using their own electrical consumption data. 
In this section, we explain in detail LSTM and how it comes to use in the forecasting, as well as FL and how it is used in our system model. 
\begin{figure}[t]
	\centering
	\includegraphics[scale=0.39]{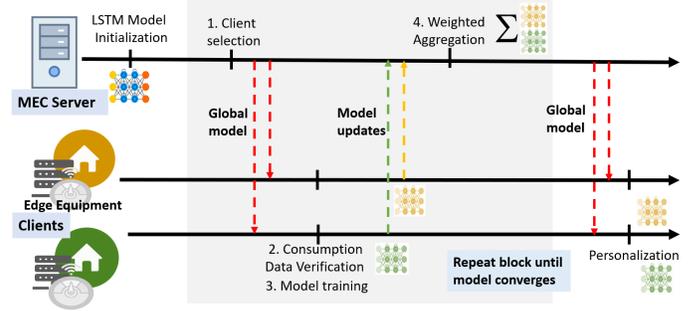}
	\caption{Network components and roles}
	\label{fig:fig_systemModel}
\end{figure}

\subsection{Time series forecasting using LSTM}
The prediction of the future electrical load in this work is achieved through the time series forecasting approach with LSTM. A time series refers to an ordered sequence of equally-spaced data points that represent the evolution of a specific variable over time. Time series forecasting is enabled through modeling the dependencies between the points of current data points and historical data, but the accuracy of the predictions relies heavily on the chosen model and the quality of historical data points.
 
LSTM is a recurrent neural network (RNN) that is fundamentally  different  from  traditional  feedforward Neural networks, and more efficient than standard RNNs.
  Sequence learning is LSTM's Forte. It is able to establish the temporal correlations between previous data points and the current circumstances, while solving vanishing and exploding gradient problems that are common in RNNs. Gradient vanishing means that  the norm of the gradient for  long-term  components  gets smaller causing weights to never change at lower layers,  while  the  gradient  exploding  refers  to the opposite event \cite{lstm1}.
  This is achieved through its key components: the memory cell that is used to remember important states in the past, and the gates that regulate the flow of information. LSTM has three gates: the input gate, the output gate and the forget gate. They learn to reset the memory cell for unimportant features during the learning process. Almost all state of the art results in sequence learning are achieved with LSTM and its variants especially  language translation and speech recognition.  In the case of residential STLF, it is expected that the LSTM network would be able to form an abstraction of some residents’ states  from  the  provided consumption profile,  maintain  the  memory  of  the  states,  and make a forecast of the future consumption based on the learnt information.

\subsection{Federated Learning}
Federated learning is a form of machine learning where most of the training process is done in a distributed way among devices referred to as clients. It was first proposed and implemented by Google on keyboards of mobile devices for next word prediction \cite{f1}.  This approach is ideal for many cases: 1) When data is privacy sensitive,
2) when data is large in size compared to model updates, 
3) highly distributed systems where the number of devices is orders of magnitude larger than nodes in a data center,
4) in supervised training when labels can be inferred directly from the user.
Federated learning has also proven to be very useful when datasets are unbalanced or non-identically distributed. 

An iteration of federated learning goes as follows : First, a subset of clients is chosen and each of them receives the current model. In our case, clients are hosted at Edge equipment in houses (e.g. smart meters). Clients that were selected compute Stochastic Gradient Descent (SGD) updates on locally-stored data, then a server aggregates the client updates to build a new global model. The new model is sent back to another subset of clients. This process is repeated until the desired prediction accuracy is reached. The operations are detailed in Algorithm 1.

In order to combine the client updates, the server uses the \textit{FederatedAveraging} algorithm \cite{f3}. First, the initial global model is initialized randomly or is pre-trained using publicly available data. In each training round $\it{r}$, the server sends a global model $w_r$ to a subset $\it{K}$ of clients who have enough data records and whose consumption load varies enough to enrich the training data. This condition was added to ensure that we have enough variation in terms of data points to give a representation of the occupants' regular consumption.  Afterward, every client $\it{k}$ in the subset uses $n_k$ examples from its local data. In our case, the volume is related to how long the smart meter has been generating data and how many of it is saved locally. The used dataset is composed of sliding windows with a predetermined number of look-back steps.
SGD is then used by each client $\it{k}$ to compute the average gradient $g_{k}$, with a learning rate $\eta$. The updated models $w_{k}$ are sent to the server to be aggregated.\\
\begin{algorithm}[h]
	\begin{algorithmic}[1]
	\State initialize the model in training round $\it{r}=0$\;	\\
	\While{$\it{r}<r_{max}$ }
		\State Select subset $\it{K}$ of clients;\;\\
		\For {client $\it{k}$ in $\it{K}$}
			\If{$\sigma(monthly load)>\it{threshold}$}
			 \State $\it{k}$ receives model $w_{r}$;\;\\
			 \State $\it{k}$ computes average gradient $g_{k}$ with SGD;\;\\
			\State $\it{k}$ updates local model \State $w_{r+1}^{k}\leftarrow w_{r}^{k}-\eta g_{k}$;\;\\
			\State $\it{k}$ sends updated model to server;\;
			\EndIf
		\EndFor
		\State server computes new global model using the equation :$w_{r+1}\leftarrow \sum_{k=0}^{K}\frac{n_{k}}{N}w_{r+1}^{k}$;\; 
		\State start next round $r\leftarrow r+1$;\; 		
	\EndWhile
	\end{algorithmic}
	\caption{Federated Averaging Algorithm. $r_{max}$ is the maximum number of rounds. $\eta$ is the learning rate and $N=\sum_{k} n_{k}$ }
\end{algorithm}

However, the centralized model may not fit all the users' electrical consumption. A  proposed solution  to  this  problem  is Personalization.  Personalization is the focus of many applications that require understanding user behaviour and adapting to it. It consists on retraining the centralized  model  using user-specific  data  to  build  a  personalized  model  for  each user. This can be achieved through retraining the model for a small number of epochs locally using exclusively the user's data \cite{f6}. 

Federated learning has fewer  privacy  risks  than  centralized server  storage,  since even  when data  are  anonymized, the users' identities are still at risk and can be discovered through reverse engineering.
The model updates sent  by each client are ephemeral and  never  stored  on  the  server; weight updates  are processed  in  memory  and  are discarded  after aggregation.
The  federated learning  procedure requires that the  individual weight uploads will not be inspected or analyzed. This is still more-secure than server training because the network and the server cannot be entrusted with fine-grained user data. Some data still have to be sent in an aggregated form for billing, but these data do not reveal many details. Techniques such as secure aggregation \cite{f5} and differential privacy\cite{f4}  are being explored to enforce trust requirements.  

\subsection{Networking Load Gain}
To evaluate the gain in network load in FL contrast to centralized training, we first define the network load $L_{sC}$ for a server $\it{s}$ in centralized training in Eq. 1 and the network load in FL $L_{sF}$ in Eq. 2. \\$S_{k-d}$ is the size of data sent by the client $\it{k}$ and $S_{m}$ is the size of the model. In the centralized training, $d_{k}$ is the number of hops between client $\it{k}$ and the server. 
\begin{equation}
L_{sC} =  \sum_{k=1}^{N} S_{k-d} \times d_{k}      
\end{equation}
\begin{equation}
L_{sF} =  S_{m} \times \sum_{r=1}^{r_{max}}\sum_{k=1}^{K}d_{k,r}    
\end{equation}
where $d_{k,r}$ is the number of hops between the client $\it{k}$ selected in round $\it{r}$ and the server, and $\it{K}$ is the number of users in each subset.\\
Using Eq.1 and Eq.2, we define the gain in networking load as follows :
\begin{equation}
G_{s} = 1 -  L_{sF} / L_{sC}    
\end{equation}

\section{Simulation and results}
\label{sec:simulation and results}

\subsection{Dataset Pre-Processing and Evaluation Method}

This research was conducted using data from Pecan Street Inc. Dataport site. Dataport contains unique, circuit-level electricity use data at one-minute to one-second intervals for approximately 800 homes in the United States, with Photovoltaics generation and Electrical Vehicles charging data for a subset of these homes \cite{database}. We chose a subset of 200 clients who have similar properties from this dataset. It is composed of the same kind of houses (detached-family homes), located in the same area  (Texas).  The dataset is composed of records  between January 1st 2019 and March 31st 2019 with a one-hour resolution data. The weather fluctuations in this period are low, so the seasonal factor can be ignored in this study.  
The data of each client is prepared to be ready for further analysis. First, we transform the data to be in a scale between 0 and 1. Then we transform the time series into sliding windows with look-backs of size 12 and a look-ahead of size 1. Finally, we  split data into train and test subsets (90\% for training and 10\% for test). We also split the clients into two groups : 180 participating in the federated learning process, and 20 are left for further evaluation for how well the model can fit non-participating clients. 

We use RMSE and MAPE to evaluate the model's performance with regard to the prediction error. RMSE allows us to quantify the error in terms of energy, while MAPE is a percentage quantifying the size of the error relative to the real value. The expressions of RMSE and MAPE are as follows: 
\begin{equation}
RMSE = \sqrt{\frac{\sum_{i=1}^{P}(y_i-\hat{y_i})^{2}}{N}}    
\end{equation}{}
\\
\begin{equation}
MAPE = {\frac{100\%}{P}}\sum_{i=1}^{P}\left | \frac{y_i-\hat{y_i}}{y_i} \right | 
\end{equation}
where $\hat{y_i}$ is the predicted value, $y_{i}$ is the actual value and $\it{P}$ is the number of predicted values.  
\subsection{Simulations setup}
The simulations were conducted on a laptop with a 2,2 GHz Intel i7 processor and 16GB of memory and NVIDIA GeForce GTX 1070 graphic card. We used Tensorflow Federated 0.4.0 with Tensorflow 1.13.1 backend.

Hyper-parameter  tuning in deep learning models is important to obtain the best forecasting performance. However, in  this  work,  we  only  focus on  evaluating the federated learning paradigm.  Previous work  shows  performance  insensitivity to combinations of some layers and layer size, as long as we use multiple layers and that the number of hidden nodes is sufficiently large \cite{s1}. It was also suggested that very deep networks are prone to under-fitting and vanishing gradients. Following these rules, the initial model hyper-parameters (e.g number of layers, and time steps to be considered) were chosen by random search on a randomly selected client's data. 
The retained model has two LSTM hidden layers composed of 200 neurons each. The loss function used is Mean squared error and the optimiser chosen is Adam. 
The model converges around the 20th epoch and thus we use close values for rounds and epochs.
\subsection{Numerical Results}

\textit{1) Evaluated scenarios:}
\newline
The different scenarios that were evaluated are summarized in Table I. As explained in the previous section, in each round, only a subset of clients train the model. We modify the number of clients in the subset selected in each round, to see the effect of larger subsets.We also vary the number of epochs of local training. In all the scenarios, the federated learning algorithm was run for 20 rounds. 
\begin{table}[h]
     	\centering{
     		\caption{Used scenarios}
     		\begin{tabular}{|c|c|c|c|c|r|r|r|r|r|}
     			
     			\hline
     			Scenarios & Clients in subset & Local Epochs      \\
     			\hline
     			 1   & 5  & 1   \\ \hline
     			 2   & 20  & 1   \\ \hline
     			 3   & 5  & 5    \\ \hline
     			 4   & 20  & 5   \\ \hline
     		\end{tabular}
     	}
     	
     	\label{tab:tab1}
     \end{table}

\textit{2) Results for global models:}
\newline
The evaluated scenarios resulted in global models that are obtained following the federated learning approach. These models are evaluated in terms of RMSE and MAPE as shown in Tables II and III. Null consumption values have been disgarded when calculating MAPE.
Table II summarizes the results for the participating clients in the different scenarios.
 In our case, the load forecast is on a granular level (single house) and on a short term (1 hour), therefore the values of MAPE achieved in Table II for various models are reasonable, and this level  of accuracy is anticipated as similar values have been reported by previous works \cite{s1,s3}. These works also report that the forecasting  accuracy  tends  to be low for short-term forecasting horizons.
 One of the most notable things we notice is that the global model fits some clients better than others when considering the fact that not all clients have similar profiles.  
 We also notice that selecting a bigger number of clients in each round is preferable, but in cases where sending updates is more expensive in terms of networking, the difference can be compensated by using more local training epochs. The results are similar when applied to the set of clients who did not participate in the training.
\vspace{0.2cm}
\begin{table}[h]
     	\centering{
     		\caption{Resulting RMSE and MAPE for global models in the considered scenarios for the 180 participating clients}
     		\begin{tabular}{|c|c|c|c|c|c|c|}
     		\hline
     			 &\multicolumn{3}{c|}{RMSE} & \multicolumn{3}{c|}{MAPE}\\ 
     			\hline
     			Scenario & Min   & Max  & Mean & Min & Max & Mean \\
     			\hline
     			1   & 0.070 & 2.652 & 0.605 & 10.65\% & 83.35\% & 41.40\%  \\ \hline
     			2   & 0.045 & 2.55 & 0.578 & 9.18\% & 87.63\% & 38.39\% \\ \hline
     			3   & 0.026 & 2.652 & 0.576 & 9.45\% & 96.84\% & 37.43\% \\ \hline
     			4   & 0.047 & 2.68 & 0.583 & 9.71\% & 93.74\% & 38.91\% \\ \hline
     			
     		\end{tabular}
     	}
     	
     	\label{tab:tab2}
     \end{table}
     
\begin{table}[h]
     	\centering{
     		\caption{Resulting RMSE and MAPE for global models in the considered scenarios for the 20 non-participant clients}
     		\begin{tabular}{|c|c|c|c|c|c|c|}
     		\hline
     			 &\multicolumn{3}{c|}{RMSE} & \multicolumn{3}{c|}{MAPE}\\ 
     			\hline
     			Scenario & Min   & Max  & Mean & Min & Max & Mean \\
     			\hline
     			1   & 0.262 & 1.024 & 0.589 & 15.82\% & 60.72\% & 44.98\%  \\ \hline
     			2   & 0.241 & 0.979 & 0.550 & 16.08\% & 55.34\% & 40.95\% \\ \hline
     			3   & 0.229 & 0.99 & 0.530 & 15.78\% & 53.98\% & 39.18\% \\ \hline
     			4   & 0.235 & 1.004 & 0.543 & 16.04\% & 56.61\% & 41.15\% \\ \hline
     		\end{tabular}
     	}
     	
     	\label{tab:tab3}
     \end{table}

\textit{3) Behaviour of personalization:}
\newline
In this section, we study the effect of personalization on the performance of the models. First we test if re-training the model locally for the participant clients gives better results. Then we apply the same thing to the set of clients who did not participate in the training. The models were retrained for 5 epochs for each client. Results for the set of clients participating in the training are summarized in Table IV and for the non-participating clients in Table V. We notice an overall improvement of most of the models. For example, the model 1 has an overall improvement of 5.07\% in terms of MAPE for the participating set of clients and of 4.78\% on the non-participating clients set. However, for some clients, the performance can not be improved despite retraining, and this, as we mentioned earlier, is related to the quality of historical data points.  Applying the models to these clients' consumption profiles results in very high MAPE, which affects the average results. These clients should be treated as outliers, nonetheless, this is beyond the scope of this study. 
\begin{table}[h]
     	\centering{
     		\caption{Resulting RMSE and MAPE after personalization over 180 clients}
     		\begin{tabular}{|c|c|c|c|c|c|c|}
     		\hline
     			 &\multicolumn{3}{c|}{RMSE} & \multicolumn{3}{c|}{MAPE}\\ 
     			\hline
     			Scenario & Min   & Max  & Mean & Min & Max & Mean \\
     			\hline
     			1   & ~0.0 & 2.47 & 0.550 & 8.13\% & 99.16\% & 36.33\%  \\ \hline
     			2   & ~0.0 & 2.47 & 0.551 & 7.89\% & 91.23\% & 36.39\% \\ \hline
     			3   & ~0.0 & 2.371 & 0.536 & 7.64\% & 88.76\% & 34.27\% \\ \hline
     	    	4   & ~0.0 & 2.375 & 0.536 & 8.00\% & 82.14\% & 34.14\% \\ \hline
     		\end{tabular}
     	}
     	
     	\label{tab:tab4}
     \end{table}
\begin{table}[h]
     	\centering{
\caption{Resulting RMSE and MAPE after personalization for 20 non-participating clients}
     		\begin{tabular}{|c|c|c|c|c|c|c|}
     		\hline
     			 &\multicolumn{3}{c|}{RMSE} & \multicolumn{3}{c|}{MAPE}\\ 
     			\hline
     			Scenario & Min   & Max  & Mean & Min & Max & Mean \\
     			\hline
     			1   & 0.232 & 0.905 & 0.516 & 18.35\% & 53.70\% & 40.20\%  \\ \hline
     			2   & 0.233 & 0.901 & 0.516 & 16.99\% & 58.68\% & 40.71\% \\ \hline
     			3   & 0.235 & 0.909 & 0.516 & 15.79\% & 54.82\% & 39.49\% \\ \hline
     			4   & 0.232 & 0.907 & 0.509 & 15.96\% & 52.96\% & 39.01\% \\ \hline
     		\end{tabular}
     	}
     
     	\label{tab:tab5}
     \end{table}
   
To illustrate the improvements on predictions using personalization, we randomly selected a client from the participant set (client 4313) and a client from the non-participant set (client 8467). We applied the global model 4 and the corresponding personalized models. The actual load profiles and the predicted profiles are shown in Fig.3 and Fig.4. Both models fit the overall behaviour of the consumption profiles. \\We conclude that we can indeed train powerful models for a population's consumption profiles using only a subset of the users forming it. For applications that have high accuracy requirements, the model can be retrained resulting in a personalized model that follows the profile's curves better, yielding more accurate predictions. Nonetheless, the predictions obtained with the global model can be a good starting point for new clients who don't have enough data for personalization. 
\begin{figure}[t]
	\centering
	\includegraphics[scale=0.68]{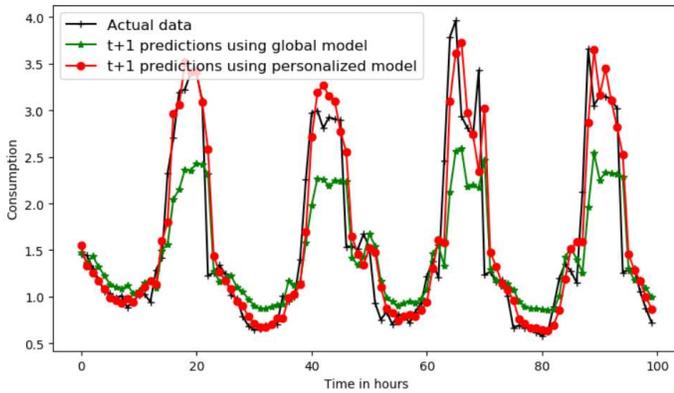}
	\caption{Predictions for next hour consumption for client 4313 who participated in training the global model 4. Local training for 5 epochs reduced RMSE from 0.55 kW to 0.388 kW.}
	\label{fig:fig_prediction}
\end{figure}
\newline
\begin{figure}[t]
	\centering
	\includegraphics[scale=0.68]{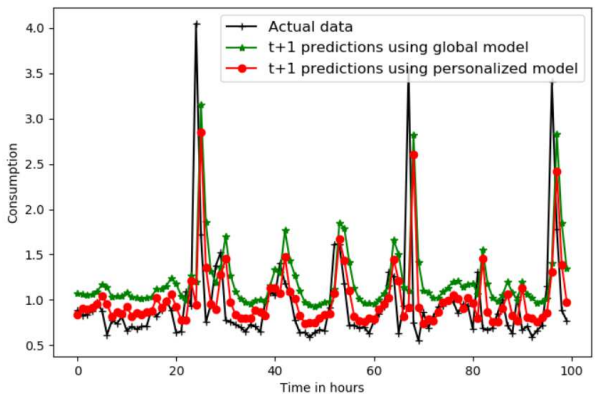}
	\caption{Predictions for next hour consumption for client 8467 who did not participate in training the global model 4. Local training for 5 epochs reduced RMSE from 0.8 kW to 0.72kW.}
	\label{fig:fig_prediction2}
\end{figure}
\captionsetup{belowskip=2pt,aboveskip=4pt}
\textit{4) Gain in network load:}
\newline
To illustrate the gain in the network load, we can consider the most basic case where the distance between all the clients and the MEC server is 1-Hop. The size of the model is 1,9Kb and the size of the used data is 16Mb. Using Eq.3, the gain in the scenarios 1 and 3 is 97\%, while scenarios 2 and 4 result in a gain of 90\%. This is a significant gain, especially when considering that the approach could be applied at the scale of a city or bigger, for example.

\section{Remarks \& future work}
\label{sec:limitations}
The feasibility of the proposed approach is dependent on the capabilities of the edge devices to perform local training.  New IoT devices have sufficient computing hardware to run complex machine learning models, but training a neural network is very likely to compromise device performance. However, some lightweight machine learning frameworks have emerged such as Tensorflow Lite \footnote{https://www.tensorflow.org/lite} which provides solid ground for future implementations. \\
The accuracy of the models, even after personalization, still varies depending on the user. To improve the results, neural networks should be coupled with other methods, such as a prior clustering of clients using criteria other than the geographical proximity. Solving the problem of outliers in this context should also be investigated.

\section{Conclusion}
\label{sec:conclusion}
Individual short-term load forecasting is a challenging task considering the stochastic nature of consumption profiles. In this paper, we proposed a system model using Edge computing and federated learning to tackle privacy and data diversity challenges related to short-term load forecasting in the smart grid.To the best of our knowledge, this represents one of the first studies of federated learning in the smart grid context.
Unlike centralized methods, in the proposed system federated learning uses edge devices to train models, hence reducing security risks to the ones related to the device only. 
We conducted experiments to evaluate the performance of both centralized and personalized models in federated settings. The simulations results show that it is a promising approach to create highly performing models with a significantly reduced networking load compared to a centralised model,  while preserving the privacy of consumption data. 


\section*{ACKNOWLEDGEMENT}
\label{sec:acknowledgement}
The authors would like to thank the Natural Sciences and Engineering Research Council of Canada, for the financial support of this research.

\bibliographystyle{unsrt}
\bibliography{./includes/references}

\begin{thebibliography}{10}

\bibitem{i1}
Elena Mocanu, Phuong~H. Nguyen, Madeleine Gibescu, and Wil~L. Kling.
\newblock Deep learning for estimating building energy consumption.
\newblock {\em Sustainable Energy, Grids and Networks}, 6:91--99, June 2016.

\bibitem{r5}
W.~Kong, Z.~Y. Dong, D.~J. Hill, F.~Luo, and Y.~Xu.
\newblock Short-{Term} {Residential} {Load} {Forecasting} {Based} on {Resident}
  {Behaviour} {Learning}.
\newblock {\em IEEE Transactions on Power Systems}, 33(1):1087--1088, January
  2018.

\bibitem{said_advanced_2013}
Dhaou Said et~al.
\newblock Advanced scheduling protocol for electric vehicle home charging with
  time-of-use pricing.
\newblock pages 6272--6276, June 2013.
\newblock ISSN: 1938-1883.

\bibitem{r2}
A.~Almalaq and J.~J. Zhang.
\newblock Evolutionary {Deep} {Learning}-{Based} {Energy} {Consumption}
  {Prediction} for {Buildings}.
\newblock {\em IEEE Access}, 7:1520--1531, 2019.

\bibitem{r3}
Salah Bouktif, Ali Fiaz, Ali Ouni, and Mohamed~Adel Serhani.
\newblock Optimal {Deep} {Learning} {LSTM} {Model} for {Electric} {Load}
  {Forecasting} using {Feature} {Selection} and {Genetic} {Algorithm}:
  {Comparison} with {Machine} {Learning} {Approaches} †.
\newblock {\em Energies}, 11(7):1636, July 2018.

\bibitem{filali_prediction-based_2019}
Abderrahime Filali et~al.
\newblock Prediction-{Based} {Switch} {Migration} {Scheduling} for {SDN} {Load}
  {Balancing}.
\newblock In {\em {ICC} 2019 - 2019 {IEEE} {International} {Conference} on
  {Communications} ({ICC})}, pages 1--6, May 2019.
\newblock ISSN: 1938-1883.

\bibitem{i4}
Yanbo Huang.
\newblock Advances in {Artificial} {Neural} {Networks} – {Methodological}
  {Development} and {Application}.
\newblock {\em Algorithms}, 2(3):973--1007, September 2009.

\bibitem{r6}
B.~Stephen, X.~Tang, P.~R. Harvey, S.~Galloway, and K.~I. Jennett.
\newblock Incorporating {Practice} {Theory} in {Sub}-{Profile} {Models} for
  {Short} {Term} {Aggregated} {Residential} {Load} {Forecasting}.
\newblock {\em IEEE Transactions on Smart Grid}, 8(4):1591--1598, July 2017.

\bibitem{r7}
H.~Shi, M.~Xu, and R.~Li.
\newblock Deep {Learning} for {Household} {Load} {Forecasting}—{A} {Novel}
  {Pooling} {Deep} {RNN}.
\newblock {\em IEEE Transactions on Smart Grid}, 9(5):5271--5280, September
  2018.

\bibitem{f3}
H.~Brendan {McMahan}, Eider Moore, Daniel Ramage, Seth Hampson, and Blaise
  Agüera~y Arcas.
\newblock Communication-efficient learning of deep networks from decentralized
  data.

\bibitem{lstm1}
S.~Hochreiter and J.~Schmidhuber.
\newblock Long short-term memory.
\newblock {\em Neural Computation}, 9(8):1735--1780, November 1997.

\bibitem{i2}
Salah Bouktif, Ali Fiaz, Ali Ouni, and Mohamed~Adel Serhani.
\newblock Optimal {Deep} {Learning} {LSTM} {Model} for {Electric} {Load}
  {Forecasting} using {Feature} {Selection} and {Genetic} {Algorithm}:
  {Comparison} with {Machine} {Learning} {Approaches} †.
\newblock {\em Energies}, 11(7):1636, July 2018.

\bibitem{i3}
{Jian Zheng}, {Cencen Xu}, {Ziang Zhang}, and {Xiaohua Li}.
\newblock Electric load forecasting in smart grids using
  {Long}-{Short}-{Term}-{Memory} based {Recurrent} {Neural} {Network}.
\newblock In {\em 2017 51st {Annual} {Conference} on {Information} {Sciences}
  and {Systems} ({CISS})}, pages 1--6, March 2017.

\bibitem{r1}
D.~L. Marino, K.~Amarasinghe, and M.~Manic.
\newblock Building energy load forecasting using {Deep} {Neural} {Networks}.
\newblock In {\em {IECON} 2016 - 42nd {Annual} {Conference} of the {IEEE}
  {Industrial} {Electronics} {Society}}, pages 7046--7051, October 2016.

\bibitem{r4}
Guangya Zhu, Tin-Tai Chow, and Norman Tse.
\newblock Short-term load forecasting coupled with weather profile generation
  methodology.
\newblock {\em Building Services Engineering Research and Technology},
  39(3):310--327, May 2018.

\bibitem{said_scheduling_2014}
Dhaou Said et~al.
\newblock Scheduling protocol with load managementfor {EV} charging.
\newblock In {\em 2014 {IEEE} {Global} {Communications} {Conference}}, pages
  362--367, December 2014.
\newblock ISSN: 1930-529X.

\bibitem{rezgui_smart_2017}
Jihene Rezgui et~al.
\newblock Smart charge scheduling for {EVs} based on two-way communication.
\newblock In {\em 2017 {IEEE} {International} {Conference} on {Communications}
  ({ICC})}, pages 1--6, May 2017.
\newblock ISSN: 1938-1883.

\bibitem{r8}
P.~Kumar, Y.~Lin, G.~Bai, A.~Paverd, J.~S. Dong, and A.~Martin.
\newblock Smart {Grid} {Metering} {Networks}: {A} {Survey} on {Security},
  {Privacy} and {Open} {Research} {Issues}.
\newblock {\em IEEE Communications Surveys Tutorials}, pages 1--1, 2019.

\bibitem{r9}
M.~Badra and S.~Zeadally.
\newblock Design and {Performance} {Analysis} of a {Virtual} {Ring}
  {Architecture} for {Smart} {Grid} {Privacy}.
\newblock {\em IEEE Transactions on Information Forensics and Security},
  9(2):321--329, February 2014.

\bibitem{r10}
Y.~Gong, Y.~Cai, Y.~Guo, and Y.~Fang.
\newblock A {Privacy}-{Preserving} {Scheme} for {Incentive}-{Based} {Demand}
  {Response} in the {Smart} {Grid}.
\newblock {\em IEEE Transactions on Smart Grid}, 7(3):1304--1313, May 2016.

\bibitem{r11}
H.~Park, H.~Kim, K.~Chun, J.~Lee, S.~Lim, and I.~Yie.
\newblock Untraceability of {Group} {Signature} {Schemes} based on {Bilinear}
  {Mapping} and {Their} {Improvement}.
\newblock In {\em Fourth {International} {Conference} on {Information}
  {Technology} ({ITNG}'07)}, pages 747--753, April 2007.

\bibitem{mec}
Quoc-Viet Pham, Fang Fang, Vu~Nguyen Ha, Mai Le, Zhiguo Ding, Long~Bao Le, and
  Won-Joo Hwang.
\newblock A {Survey} of {Multi}-{Access} {Edge} {Computing} in 5g and {Beyond}:
  {Fundamentals}, {Technology} {Integration}, and {State}-of-the-{Art}.
\newblock {\em arXiv:1906.08452 [cs, math]}, June 2019.
\newblock arXiv: 1906.08452.

\bibitem{f1}
Andrew Hard, Kanishka Rao, Rajiv Mathews, Swaroop Ramaswamy, Françoise
  Beaufays, Sean Augenstein, Hubert Eichner, Chloé Kiddon, and Daniel Ramage.
\newblock Federated learning for mobile keyboard prediction.

\bibitem{f6}
Khe~Chai Sim, Petr Zadrazil, and Françoise Beaufays.
\newblock An {Investigation} into {On}-{Device} {Personalization} of
  {End}-to-{End} {Automatic} {Speech} {Recognition} {Models}.
\newblock In {\em Interspeech 2019}, pages 774--778. ISCA, September 2019.

\bibitem{f5}
Keith Bonawitz, Vladimir Ivanov, Ben Kreuter, Antonio Marcedone, H.~Brendan
  McMahan, Sarvar Patel, Daniel Ramage, Aaron Segal, and Karn Seth.
\newblock Practical {Secure} {Aggregation} for {Federated} {Learning} on
  {User}-{Held} {Data}.
\newblock {\em arXiv:1611.04482 [cs, stat]}, November 2016.
\newblock arXiv: 1611.04482.

\bibitem{f4}
Robin~C. Geyer, Tassilo Klein, and Moin Nabi.
\newblock Differentially {Private} {Federated} {Learning}: {A} {Client} {Level}
  {Perspective}.
\newblock {\em arXiv:1712.07557 [cs, stat]}, December 2017.
\newblock arXiv: 1712.07557.

\bibitem{database}
Pecan street inc. dataport 2019 [online] https://dataport.pecanstreet.org/.

\bibitem{s1}
W.~Kong, Z.~Y. Dong, Y.~Jia, D.~J. Hill, Y.~Xu, and Y.~Zhang.
\newblock Short-{Term} {Residential} {Load} {Forecasting} {Based} on {LSTM}
  {Recurrent} {Neural} {Network}.
\newblock {\em IEEE Transactions on Smart Grid}, 10(1):841--851, January 2019.

\bibitem{s3}
Matthew Rowe, Timur Yunusov, Stephen Haben, William Holderbaum, and Ben Potter.
\newblock The {Real}-{Time} {Optimisation} of {DNO} {Owned} {Storage} {Devices}
  on the {LV} {Network} for {Peak} {Reduction}.
\newblock {\em Energies}, 7(6):3537--3560, June 2014.

\end{thebibliography}


\end{document}